\documentclass[aps,pra,twocolumn,showpacs,preprintnumbers,amsmath,amssymb,superscriptaddress,10pt,nofootinbib]{revtex4-1}

\usepackage{amsmath,amssymb}
\usepackage{bm}
\usepackage[latin1]{inputenc} 
\usepackage{dcolumn}
\usepackage{graphicx}
\usepackage{SIunits}
\usepackage{ae}
\usepackage{color} 

\newcommand{\diff}[0]{\text{d}}

\newcommand{\re}{\mathrm{Re}\,}
\newcommand{\im}{\mathrm{Im}\,}

\newcommand{\be}{\begin{equation}}
\newcommand{\ee}{\end{equation}}
\newcommand{\ba}{\begin{align}}
\newcommand{\ea}{\end{align}}

\newcommand{\omegaobs}{\omega_0}

\newcommand{\kz}{k_{z}}
\newcommand{\kzz}{k_{z}'}

\newcommand{\Ekx}{E(k_x,z,\omega)}
\newcommand{\eps}{\epsilon_1}
\newcommand{\epss}{\epsilon_2}
\newcommand{\thetac}{\theta_{\text{c}}}
\newcommand{\kpi}{k_{\text{p}j}}

\begin{document}

\title{Laplace-Fourier analysis and instabilities of a gainy slab}

\author{Hans Olaf Hågenvik}
\author{Johannes Skaar}
\affiliation{Department of Electronics and Telecommunications, Norwegian University of Science and Technology, NO-7491 Trondheim, Norway}
\email{hans.hagenvik@iet.ntnu.no}

\date{\today}

\begin{abstract}
The idealization of monochromatic plane waves leads to considerable simplifications in the analysis of electromagnetic systems. However, for active systems this idealization may be dangerous due to the presence of growing waves. Here we consider a gainy slab, and use a realistic incident beam, which is both causal and has finite width. This clarifies some apparent paradoxes arising from earlier analyses of this setup. In general it turns out to be necessary to involve complex frequencies $\omega$ and/or complex transversal wavenumbers $k_x$. Simultaneously real $\omega$ and $k_x$ cannot describe amplified waves in a slab which is infinite in the transversal direction. We also show that the only possibility to have an absolute instability for a finite width beam, is if a normally incident plane wave would experience an instability. 
\end{abstract}

\maketitle

\section{Introduction}
Physicists love idealizations and simplifications, to understand the main principles of physical systems without being disturbed by unnecessary complications or degrees of freedom. The idealizations may make it possible to get analytical results for complicated systems which otherwise call for advanced numerical methods. Without the idealizations, the physicist's understanding and intuition would suffer, and the ability to master a large parameter space would be very limited.

However, sometimes the involved idealizations may be conflicting or give unphysical results. When considering the optical or electromagnetic response of a slab, it is convenient to let the slab be infinite in the transversal direction, to eliminate complications from reflection at the transversal boundaries. At the same time, it is customary to let the incident wave be monochromatic and plane, extending infinitely in time and transversal space. These simplifications have been proved very useful for passive media, and have been used in a number of settings (see e.g. \cite{saleh,orfanidis,pendry2000}).

Even for gainy slabs, it is quite common to do a monochromatic plane wave analysis \cite{callary76,*kolokolov75,*lukosz76,*nazarov07,*nazarov08,mansuripur14}. Clearly, such an analysis is appealing due to its simplicity, and may give correct results for certain situations. However, it has been argued that this analysis is dangerous in general, and may lead to unphysical results \cite{skaar06,nistad08,pukhov,perezmolina08}: Real physics happens in the time- and spatial domain, and to justify the monochromatic plane wave analysis, Fourier transforms of the fields with respect to time $t$ and the transversal coordinate $x$ must exist.

In this paper we consider a slab which extends infinitely in the transversal $x$-direction; however we vary the description of the source: In Sec. \ref{naive} the incident beam is described by monochromatic, plane waves. This gives peculiar results which we later argue are incorrect. In particular, the refraction of the beam inside the slab changes drastically (from positive to negative), subject to a small increase of the incident angle. This happens despite the fact that the slab is made of a conventional, dielectric, weak gain medium.

It is therefore natural to consider a more realistic source. In Sec. \ref{causal} we consider a causal, plane wave source, starting at some time $t=0$. Due to the possibility of instabilities, we use the inverse Laplace transform to express the time-domain fields from frequency components. This analysis, which has been discussed in the literature previously \cite{skaar06,nistad08,pukhov,perezmolina08}, gives physically reasonable results. However, it has the disadvantage of predicting absolute instabilities related to the infinite extent of the plane wave in the $x$-direction. These absolute instabilities are nonexistent in real experiments where the transversal dimension of the source is limited \cite{degrasse59,*geusic62,*koester64,*koester66,*kogan72,*becker}. 

Thus we finally describe the source realistically as a causal beam of finite width (Sec. \ref{fw}). This analysis, although somewhat more complicated, gives the intuitive results that we expect from previous experiments. Moreover, by deforming the paths of the inverse integral transforms, it turns out that the system can often be described monochromatically, provided we allow the transversal wavenumbers $k_x$ to be complex in the superposition. We derive an explicit expression for the difference between the solution from the monochromatic plane wave analysis and the proper analysis, and demonstrate that it is the neglect of this term that lead to the incorrect results in Sec. \ref{naive}. This fact is also independently verified by finite-difference-time-domain (FDTD) simulations  (Sec. \ref{fdtd}).

Our results are quite intuitive: A gain medium leads to growing fields. When a finite width beam has oblique incidence, we expect the fields in the slab to grow along the transversal $x$ direction. Therefore, to describe the fields in $(\omega,k_x)$-space, it is necessary to let either the angular frequency $\omega$ or the transversal wavenumber $k_x$ be complex. Modes with simultaneously real $\omega$ and $k_x$ cannot describe amplification in the $x$-direction in an unbounded slab.

When considering growing waves, an instability is classified as absolute if the fields grow unlimited with time even at a fixed point in space \cite{sturrock,briggs}. Otherwise, an instability is convective. The movement of singularities of the field as we deform the complex integration paths in $\omega$ and $k_x$ space \cite{briggs,kolokolov99,hagenvik14}, is used to classify the possible instabilities in the slab. It turns out that the only possibility of having an absolute instability, is if the mode $k_x=0$ is unstable. That is, to find out whether the slab supports absolute instabilities for finite width beams, we only need to check plane waves at normal incidence, and determine if multiple reflections inside the slab are being amplified enough to compensate for the transmission out of the boundaries.

\begin{figure}
	\centering
	\includegraphics[width=5cm]{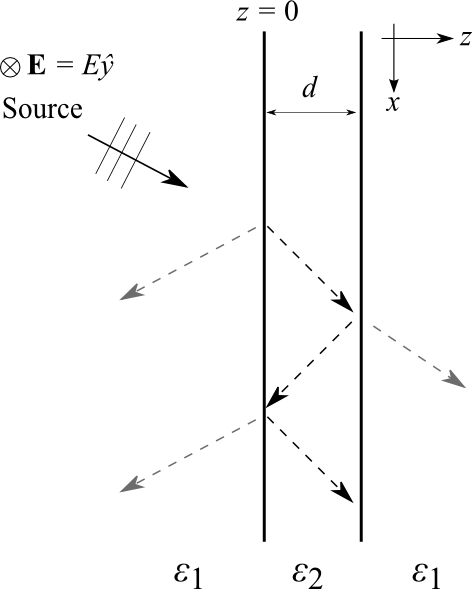}
	\caption{The setup consists of a slab with permittivity $\epss$, surrounded by a passive medium $\eps$. The dimensions of the slab are infinite in the $x$ and $y$ directions, with thickness $d$ in the $z$ direction. The incident wave is polarized along the $y$-direction.}
	\label{fig:setup}
\end{figure}

The setup is a slab of permittivity $\epss$ surrounded by a passive medium $\eps$. Both media are linear, time-invariant, isotropic, homogeneous, and without spatial dispersion. For simplicity they are assumed nonmagnetic. The slab is infinite in the $x$ and $y$ directions, and has thickness $d$ in the $z$ direction, see Fig. \ref{fig:setup}. The incident wave is polarized with its electric field along the $y$-direction. 

The linearity assumption needs some clarification. In practice any gain medium is only linear for sufficiently small fields, due to gain saturation. Thus the linear analysis is accurate in the absence of instabilities, when the incident beam is sufficiently weak. It is also useful to determine the behavior of the system before the instabilities have become too large.

\section{Monochromatic plane wave analysis}\label{naive}
In this section we assume monochromatic fields of the form $e^{-i\omega t}$, where $\omega$ is real. To compute the electric field, we apply a straightforward plane wave expansion method, considering the incident beam to be gaussian. 

By solving Maxwell's equations in the frequency domain, and applying the boundary conditions for the electric and magnetic fields, the electric field amplitude of each plane wave component $k_x$ is found to be
\begin{equation}\label{Ekx}
	\Ekx = \begin{cases}
		 e^{i\kz z} + R e^{-i\kz z}, & z < 0 \\
		 S^{+} e^{i\kzz z} + S^{-} e^{-i\kzz z}, & 0 \le z \le d \\
		 T e^{i\kz (z-d)}, & z > d.
	\end{cases}
\end{equation}
The total reflection coefficient $R$, the amplitudes in the slab $S^\pm$ and the transmission coefficient $T$ are functions of $(k_x,\omega)$ through
\begin{subequations}\label{RST}
	\begin{align}
		R &= \frac{(\kz^2 - \kzz^2) [1 - \exp(2i\kzz d)]}{(\kz + \kzz)^2 - (\kz - \kzz)^2\exp(2i\kzz d)} \\
		S^{+} &=  \frac{2\kz(\kz+\kzz)}{(\kz + \kzz)^2 - (\kz - \kzz)^2\exp(2i\kzz d)}\\
		S^{-} &= \frac{2\kz(\kz-\kzz)}{(\kz - \kzz)^2 - (\kz + \kzz)^2\exp(-2i\kzz d)}\\
		T &= \frac{4\kz\kzz\exp(i\kzz d)}{(\kz + \kzz)^2 - (\kz - \kzz)^2\exp(2i\kzz d)}. \label{RSTd}
	\end{align}
\end{subequations}
Here $\kz = \sqrt{\eps\omega^2/c^2 - k_x^2}$ and $\kzz = \sqrt{\epss\omega^2/c^2 - k_x^2}$ are the longitudinal wave number outside and inside the slab, respectively, and $c$ is the vacuum light velocity. It is worth noting that the expressions \eqref{Ekx} do not depend on which sign we choose for $\kzz$. The physical electric field is given as a superposition of all plane wave components $k_x$, by an inverse Fourier transform. The weighting of each component $k_x$ is given by the distribution of the incident beam. For a gaussian beam the plane wave spectrum is gaussian, centered around the main plane wave component.

For concreteness, we consider a slab with permittivity given as an inverted Lorentzian,
\begin{equation}\label{inv_lorentz}
	\epss(\omega) = 1 - \frac{F\omega_0^2}{\omega_0^2 - \omega^2 - i\Gamma\omega},
\end{equation}
where $\omega_0$ is the resonance frequency, $F>0$ (gain) and the gain is weak ($F\omega_0/\Gamma \ll 1$). In all calculations here and below, we let $F=0.001$ and $\Gamma=0.1 \omega_0$. Moreover, $\epsilon_1=2.25$, and the thickness $d$ of the slab is given by $\omega_0 d/c=100$. For the monochromatic calculations in this section, the observation frequency is set equal to the resonance frequency $\omega_0$, at which $\epss = 1 - 0.01i$. The width $\sigma$ of the source is given by $\omega_0\sigma/c=35$.

\begin{figure}
	\centering
	\includegraphics[width=8cm]{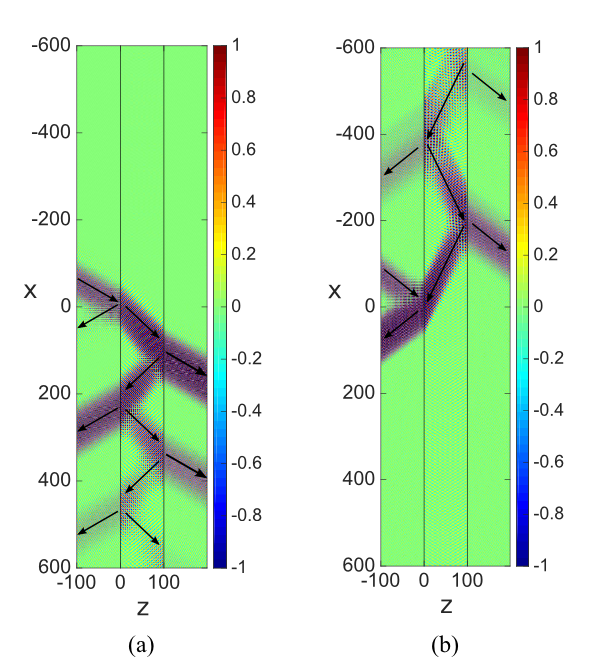}
	\caption{A gaussian beam incident to a gainy slab at incident angles of a) $30$ degrees and b) $37$ degrees. A small change of the incident angle cause the refraction inside the slab to go from positive to negative. The arrows indicate the direction of the phase velocity.}
	\label{fig:wrong}
\end{figure}

Figures \ref{fig:wrong}a and b show the resulting field, for incident angles $\theta=30$ and $\theta=37$ degrees, respectively. 
A rather peculiar observation is made: As the incident angle is altered from $30$ to $37$ degrees, the beam goes from being positively to negatively refracted inside the slab. Here ``negative refraction'' refers to the direction of the beam. As the slab consists of a weak gain medium with low dispersion, this result does not seem to be realistic.

To investigate this peculiarity further, calculations were performed for incident angles from $0$ to $90$ degrees. At small incident angles the transmitted electric field is found to increase with increasing $\theta$. Intuitively, this makes sense, as the propagation length inside the slab increases. However, for incident angles larger than $30$ degrees the field does not increase any more, but instead gradually goes over to be negatively refracted. This negatively refracted beam decreases in magnitude with distance from the entrance of the incident beam.  

As will become apparent from the next sections, these results are incorrect, and arise from the assumption that the fields are monochromatic \emph{and} built up of real plane wave components. This excludes the possibility of describing a wave that grows exponentially as $x\to\pm\infty$.

\section{A causal source}\label{causal}
In an actual experiment the source has to be turned on at some time $t=0$. Due to this abrupt onset, all frequencies will be excited to some extent. To determine the correct response of the slab we may therefore consider the physical electric field $E(x,z,t)$ as a superposition of frequency components $e^{-i\omega t}$. Since the slab is active, the system may contain instabilities, resulting from the multiple reflections inside the slab being amplified enough to compensate for the transmission out of the boundaries. In some cases the electric field is thus not Fourier transformable wrt. $t$, and interpretation at real frequencies is not meaningful. To account for these possible instabilities, the Laplace transform may be used to decompose the signal into (possibly) complex frequency components $e^{-i\omega t}$ \cite{skaar06}. 

To avoid making the analysis unnecessarily complicated we first consider only a single plane wave component $k_x$. Such an analysis of the response of an active slab to a causal excitation, where each plane wave component is treated separately, has been done in \cite{nistad08,pukhov}. For concreteness, we consider a source with time dependency
\begin{equation}\label{vt}
	v(t) = H(t) e^{-i\omegaobs t},
\end{equation}
i.e. a harmonic oscillation at frequency $\omegaobs$ turned on at $t=0$ by the Heaviside function $H(t)$. The Laplace transform of this function is
\begin{equation}
	V(\omega) = \frac{i}{\omega-\omegaobs},
\end{equation}
which constitutes the frequency distribution of the source. The electric field in the time-domain is given by the inverse Laplace transform
\begin{equation}\label{Epw}
	E(x,z,t) = \frac{1}{2\pi} \int_{i\gamma - \infty}^{i\gamma + \infty} \diff{\omega}V(\omega) \Ekx e^{ik_xx-i\omega t},
\end{equation}
where $\Ekx$ is given by \eqref{Ekx}. The transform is performed along the line $\im{\omega}=\gamma$, where $\gamma$ is some positive number larger than the maximal growth rate of the field with time. The integration path can be closed along an infinite semicircle in the lower half-plane. The residue theorem then gives the field as
\begin{equation}\label{Ew}
	E(x,z,t)=E(k_x,z,\omegaobs)e^{ik_xx-i\omegaobs t} + \text{rem}(t).
\end{equation}
Here the first term originates from the pole at $\omegaobs$, and the term rem$(t)$ is a result of all other singularities (cuts, poles etc.) of the integrand inside the closed integration contour. If all singularities of $E(k_x,z,\omega)$ are located in the lower half-plane $\im{\omega}< 0$ (for the given $k_x$) these transients will decay with time. Then, for sufficiently large times, the field is given by the first term in \eqref{Ew}.

If, on the other hand, the field has poles in the upper half-plane, we may still close the integration contour along a semicircle in the lower half-plane. However, now the term rem$(t)$ will contain terms with frequencies corresponding to the poles in the upper half-plane. These will grow with time, i.e. we have an absolute instability.

In the calculations in the previous section, $E(k_x,z,\omega)$ has poles in the upper half-plane for $k_x$ corresponding to incidence angles larger than $\thetac = 25.3$ degrees. We may therefore not assume the field to be monochromatic for plane waves incident at $\theta > \thetac$, assuming real $k_x$ in the slab. This explains the peculiar refraction in Fig. \ref{fig:wrong}b.

A perfectly plane wave is present \emph{everywhere}, and thus enters the slab at infinitely large $|x|$. Plane wave components experiencing net round trip gain as they propagate inside the slab, will therefore grow unlimited with time at any fixed point. In fact, even for a thin slab made of a weak gain medium, these absolute instabilities may occur for sufficiently large $\theta$. The instabilities are not very physical, as they are caused by the assumption that the incident wave has infinite width.

Experiments on amplifying waveguides do not suffer from instabilities, even for wave propagation at large $\theta$ (or small angles with the waveguide axis) \cite{degrasse59,*geusic62,*koester64,*koester66,*kogan72,*becker}. This is because in an experimental setup the waveguide has finite length and the source does not extend infinitely along the waveguide axis. Absolute instabilities due to propagation over infinite distances inside the waveguide will therefore not be present. To determine whether an unbounded slab, such as that in Fig. \ref{fig:setup}, supports absolute instabilities when the source is causal and has finite width, we must consider a double Laplace-Fourier transform wrt. $\omega$ and $k_x$. This is done in the next section.
 
\section{A causal source of finite width}\label{fw}
For a causal source of finite width, we find that the physical electric field is given by
\begin{align}\label{Exzt}
	E(x,z,t) &= \frac{1}{(2\pi)^2}\int_{i\gamma-\infty}^{i\gamma+\infty}\diff{\omega} V(\omega)e^{-i\omega t}\cdot \\ &  \int_{-\infty}^{\infty}\diff{k_x}U(k_x)E(k_x,z,\omega) e^{ik_xx}, \nonumber
\end{align}
by superposition of plane waves of the form \eqref{Epw}. Here $E(k_x,z,\omega)$ is given by \eqref{Ekx}, and $U(k_x)$ and $V(\omega)$ are the plane wave and frequency spectrum of the source, respectively. Note that if the source has finite width, its Fourier transform $U(k_x)$ is an analytic function (entire function).

It is reasonable to expect that for a sufficiently weak gain medium, or sufficiently small slab thickness, there will be no absolute instabilities. If this is the case, it is not necessary to describe the field using complex frequencies with $\im{\omega}>0$. There may however be convective instabilities, i.e. amplification of the field in the $\pm x$-directions. This suggests that in certain cases interpretation of the fields at real frequencies should be meaningful, provided complex wavenumbers $k_x$ are included in the superposition.

\begin{figure*}
	\centering
	\includegraphics[width=16cm]{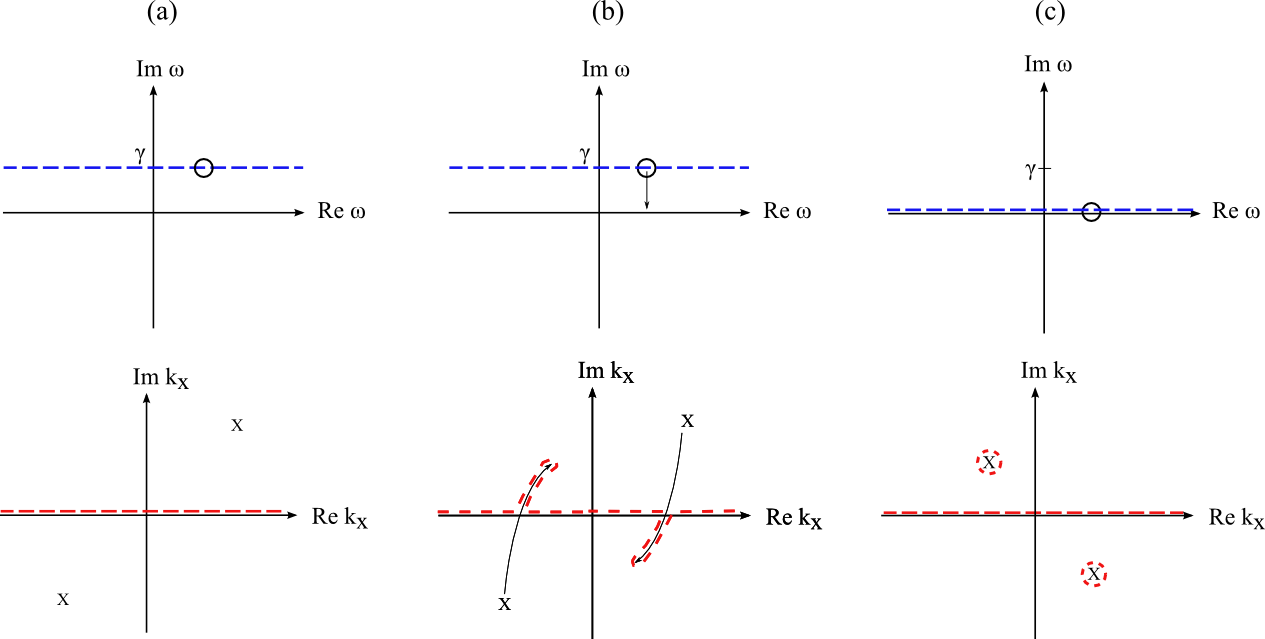}
	\caption{(a) The original integration paths in Eq. \eqref{Exzt}. In the lower figure the X's indicate poles of $E(k_x,z,\omega)$, for the $\omega$ encircled in the upper figure. (b) The poles will move as indicated by arrows in the lower figure, as we reduce $\im{\omega}$ from $\gamma$ to $0$ for the encircled $\omega$. We deform the $k_x$ integration path for this $\omega$ to allow this movement. (c) Provided such a deformation is possible for all $\omega$ along the integration path in the upper figure of (a), the inverse Laplace transform can be taken along the real axis. The lower figure shows the deformed inverse Fourier transform integration path wrt. $k_x$ in the complex $k_x$ domain, for the encircled $\omega$ in the upper figure.}
	\label{fig:deformation}
\end{figure*}

Formally, we will now employ the deformation method developed in Ref. \cite{briggs}, in which the $\omega$-integration in \eqref{Exzt} is moved down to the real axis, at the expense of deforming the $k_x$-integration into the complex plane (Fig. \ref{fig:deformation}). To this end we need to know the singularities of the field $E(k_x,z,\omega)$, which can be found by considering the denominator in \eqref{RST}. In Sec. \ref{causal} we noted that even for a weak gain medium, the field has poles for $\im{\omega}>0$ for some real $k_x$. These poles suggest that the Laplace contour cannot straightforwardly be moved down to the real axis.

However, these singularities may often be avoided, by deforming the integration path in the complex $k_x$ domain. For a given $\omega$, the field will have poles in the complex $k_x$ plane, for example as indicated in the lower part of Fig. \ref{fig:deformation}a. As we move $\omega$ along the arrow in the upper part of Fig. \ref{fig:deformation}b, the poles will e.g. move as indicated by arrows in the lower part. For frequencies along the arrow, we now deform the integration path in the complex $k_x$ domain to avoid these pole trajectories. Now, for $k_x$'s along this deformed path, there will be no poles for $\omega$'s along the arrow, and we may move the encircled segment of the $\omega$-integration path down to the real $\omega$-axis.

To be able to move the encircled $\omega$ in the upper figure of \ref{fig:deformation}a we required that the integration path wrt. $k_x$ could be deformed around the pole trajectories. We will later show that this is equivalent to requiring that there are no absolute instabilities present. If this is the case, we may reduce $\im{\omega}$ from $\gamma$ to $0$ for each segment of the inverse Laplace contour. Note that the deformation in the complex $k_x$ plane in general is different for each $\omega$-segment. Equation \eqref{Exzt} can thus be rewritten to
\begin{align}\label{Exzt_kappa}
	 E(x,z,t) &= \frac{1}{(2\pi)^2}\int_{-\infty}^{\infty} \diff{\omega}V(\omega)e^{-i\omega t}\cdot \\ &  \int_{\kappa(\omega)}\diff{k_x}U(k_x)E(k_x,z,\omega) e^{ik_xx} \nonumber
\end{align}
where $\kappa(\omega)$ is the deformed integration contour in the complex $k_x$ plane for each $\omega$. The paths $\kappa(\omega)$ can be written as a sum of integration along the real $k_x$-axis, plus detours around the pole trajectories. The detour integrals may be simplified into contour integrals around the poles (lower part of Fig. \ref{fig:deformation}c), by noting that the integrations "up and down" cancel. The integrals around the poles are calculated using the residue theorem. For some $\omega$'s the poles never cross the real axis, and no deformation is needed. For other $\omega$'s, several poles may cross the axis. In general we get 
\begin{align}\label{Edef}
	E(x,z,t)  = & \frac{1}{2\pi}\int_{-\infty}^{\infty} \diff{\omega}V(\omega) e^{-i\omega t}\cdot \\  
	( &\frac{1}{2\pi} \int_{-\infty}^{\infty} \diff{k_x}U(k_x)E(k_x,z,\omega) e^{ik_xx} \nonumber\\ 
	& + i\sum_{\text{poles } j} \pm \underset{k_x=\kpi}{\operatorname{Res}} U(k_x)E(k_x,z,\omega) e^{ik_xx} )   \nonumber,
\end{align}
where $\kpi$ is the $j$th pole which has crossed the real $k_x$-axis. The sign in front of each residue term is positive for anti-clockwise and negative for clockwise integration contours. 

We have thus found that in the absence of absolute instabilities, \eqref{Exzt} may be rewritten into \eqref{Edef}. For the excitation \eqref{vt}, \eqref{Edef} may be evaluated by closing the $\omega$-integration contour along a semicircle in the lower half-plane. For sufficiently large times, when all transients have died out, the electric field is given by
\begin{align}\label{Exzt_mc}
	E(x,z,t) &=  
	\frac{1}{2\pi} \int_{-\infty}^{\infty} \diff{k_x}U(k_x)E(k_x,z,\omegaobs) e^{ik_xx-i\omegaobs t} \nonumber\\ 
	& + i\sum_{\text{poles } j} \pm \underset{k_x=\kpi}{\operatorname{Res}} U(k_x)E(k_x,z,\omegaobs) e^{ik_xx-i\omegaobs t}.  
\end{align}
In \eqref{Exzt_mc} the first term describes the method in Sec. \ref{naive}, so the difference is the sum of residues. This sum goes over poles that have crossed the real $k_x$-axis as $\im\omega$ was reduced down to 0. These complex plane wave components will cause an exponential growth in the $\pm x$-directions, and thereby describe convective instabilities. 

\begin{figure}
	\centering
	\includegraphics[width=8cm]{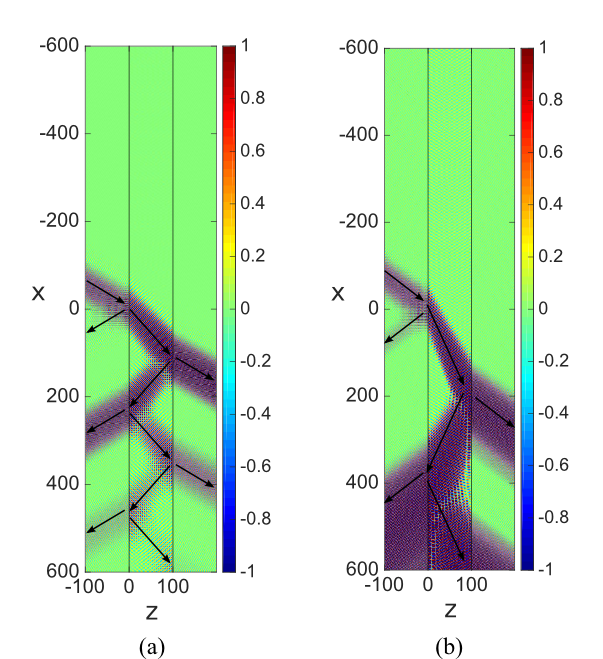}
	\caption{The electric field for the same situation as in Fig. \ref{fig:wrong}, but where the summation over residues (the second term in \eqref{Exzt_mc}) has been added to the field. The arrows indicate the direction of the phase velocity.}
	\label{fig:correct}
\end{figure}

The pole trajectories may be found numerically, by considering $|T(k_x,\omega)|$ in the complex $k_x$ plane for $\omega$ along the arrow in the upper part of Fig. \ref{fig:deformation}b. For the setup in Sec. \ref{naive}, a total of $34$ ($17$ for $\re{k_x}>0$ and $17$ for $\re{k_x}<0$) poles cross the real $k_x$-axis as we reduce $\im{\omega}$ from $\gamma$ to $0$ for $\re{\omega}=\omegaobs$. Figure \ref{fig:correct} shows the resulting field when the sum over residues has been added to the field from Fig. \ref{fig:wrong}.

The beam propagation in Figs. \ref{fig:correct}b and \ref{fig:wrong}b are fundamentally different. The summation over  residues has canceled out the seemingly negative refracted beam in \ref{fig:wrong}b, and the field instead propagates in the $+x$-direction, growing in amplitude. The refraction pattern in Figs. \ref{fig:wrong}a and \ref{fig:correct}a are virtually identical, so the summation over residues did not affect the field significantly in this case. This happens despite the fact that the same observation frequency is used, so the summation is over the same residues in both cases. The explanation for this is that the exponentially growing residue terms are excited differently, due to different angles of incidence. The poles are located in the region of $\re k_x$'s corresponding to angles of incidence between $34.3$ and $41.7$ degrees.  

In the deformation of the integration paths in \eqref{Exzt} to the ones in \eqref{Edef}, we required that it is possible to deform the inverse Fourier transform path wrt. $k_x$, to avoid the poles crossing the real $k_x$-axis as we reduce $\im{\omega}$ from $\gamma$ towards $0$. Such a deformation is impossible if two poles in the complex $k_x$-plane collide in such a way that the integration path gets ``stuck'' (see Fig. \ref{fig:crash}). It is argued in Appendix \ref{app:absinstab} that $k_x=0$ is the only plane wave component at which this can happen. If two poles collide at $k_x=0$ for some $\im{\omega}>0$, we cannot move this $\omega$ further down, and the inverse Laplace transform has to contain complex frequencies with $\im{\omega}>0$. The field will therefore grow with time, even at a fixed point (absolute instability).

\begin{figure}
	\centering
	\includegraphics[width=6cm]{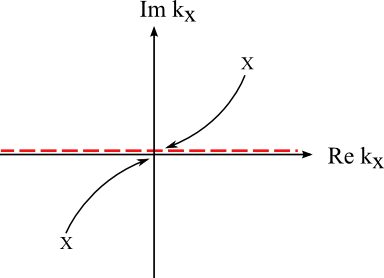}
	\caption{If two poles collide at $k_x=0$ the integration path gets ``stuck'', and we cannot move the $\omega$ integration path further down to the real axis. This means that the field necessarily contains complex frequency components with $\im{\omega}>0$, so we have an absolute instability.}
	\label{fig:crash}
\end{figure}

It is perhaps not entirely obvious that the Fourier transform in $x$ exists, provided the Laplace transform in $t$ is used. To justify \eqref{Exzt}, we can use one of the following two approaches: Either we just assume the existence of the Fourier-Laplace transform, and verify that the solution in the time-spatial domain is consistent with the assumptions. Or we consider the transforms in the opposite order. By causality, $E(x,z,t)$ has finite support as a function of $x$, for any fixed $t$. Thus it is Fourier transformable. Interchanging the order of integration can be done in light of Fubini's theorem\cite{mcdonaldweiss}, which places conditions on the source $U(k_x)V(\omega)$.

\section{Fdtd simulations}\label{fdtd}
\begin{figure}
	\centering
	\includegraphics[width=7.6cm]{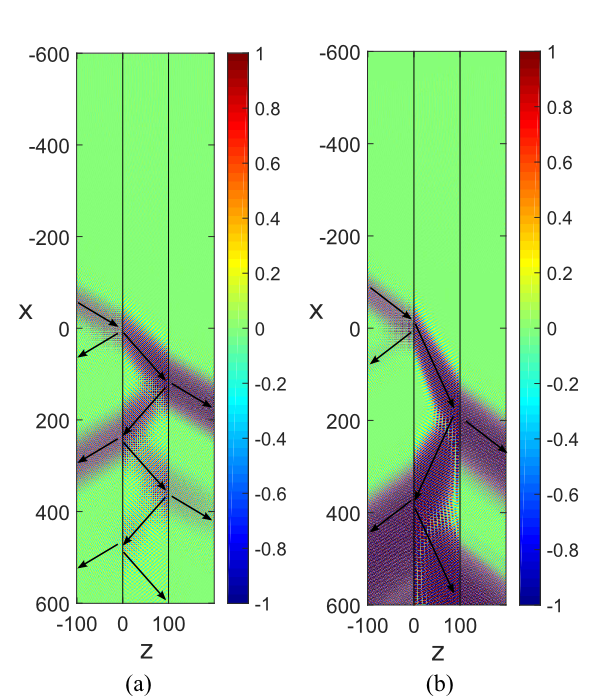}
	\caption{FDTD simulations of the same setup as in Fig. \ref{fig:wrong}. The angles of incidence are (a) $30$ and (b) $37$ degrees. The arrows indicate the direction of the phase velocity and the beam propagation.}
	\label{fig:fdtd}
\end{figure}

The analysis and calculations so far have been done in the frequency domain, as Maxwell's equations are easy to solve there. It is however crucial to remember that actual physics happens in spatial-time domain. It is therefore natural to perform finite-difference-time-domain (FDTD) simulations, as an independent verification of \eqref{Exzt_mc}. Figure \ref{fig:fdtd} shows the resulting field distribution from FDTD simulations of the same situation as in Figs. \ref{fig:wrong} and \ref{fig:correct}. Clearly, the FDTD simulations agree with the method in Sec. \ref{fw}. This supports the conclusion that the standard monochromatic plane wave solution is generally not applicable when analyzing gain media.

The simulations are based on an implementation of a FDTD method \cite{yee66} for dispersive Lorentzian media \cite{luebbers90}. The dispersion \eqref{inv_lorentz} is therefore taken into account. The simulation domain was chosen large, so that the fields never reached the boundaries of the domain during the duration of the numerical experiment.

\section{Discussion and conclusion}
We have argued that the standard description of the electric field as a superposition of real frequencies $\omega$ \emph{and} real wavenumbers $k_x$, is not very useful when analyzing media with gain. For the situation with an infinite slab, such a description requires the electric field to be simultaneously Fourier transformable wrt. $t$ and $x$. This is generally not the case, due to amplification in the slab. Therefore, $k_x$ and/or $\omega$ must be complex in the integrals. In Sec. \ref{fw} we derived an expression for the electric field in a gainy slab, by assuming a causal, incident beam of finite width. Provided there are no absolute instabilities present, interpretation of the field at a real observation frequency $\omegaobs$ is meaningful. From the explicit expression of the field \eqref{Exzt_mc}, we find that a sum of terms with complex wavenumbers $k_x$ must be added to the conventional Fourier integral in $k_x$. This fact is also independently verified using FDTD simulations.

From \eqref{Exzt_mc} we may further try to approach the plane wave limit, by letting $\sigma\to\infty$. Any physical source will have finite support, meaning the excitation is identical to $0$ for sufficiently large $x$. It can be shown that for any $u(x)$ with finite support, its Fourier transform $U(k_x)$ will diverge for complex $k_x$, when $\sigma\to\infty$. This means that the summation over residues in \eqref{Exzt_mc} will lead to diverging field in this limit. In the monochromatic limit, the plane wave limit $\sigma\to\infty$ is thus not meaningful. The physical explanation for this is that in the limit $\sigma\to\infty$, the field from the source propagates an infinite distance along the $x$-axis before reaching a given point, thus picking up an infinite amount of gain. If there are no poles to sum over in \eqref{Exzt_mc} (as always will be the case for passive media), the limit $\sigma\to\infty$ exists, and the standard solution $E(x,z,t) = E(k_x,z,\omegaobs)e^{ik_xx - i\omegaobs t}$ is obtained.

For a realistic incident beam, which is both causal and has finite width, we prove that absolute instabilities only occur if the $k_x=0$ mode has poles in the upper half-plane $\im\omega>0$. In other words, to check if the slab supports absolute instabilities, it suffices to consider a normally incident plane wave, and determine if it leads to a diverging field. When the $k_x=0$ mode experiences an instability, the slab will support absolute instabilities even for oblique incident beams of finite widths, as a result of their Fourier spectrum containing the $k_x=0$ mode.

Given that the refraction pattern in Fig. \ref{fig:wrong}b is incorrect, one may ask if a similar pattern could arise in a more complicated medium. From \eqref{Exzt_mc} it is seen that the field distribution would in fact be as in Fig. \ref{fig:wrong}b if there were no poles to sum over. In other words, if we can find a medium $\epsilon(\omega)$, with $\epsilon(\omegaobs) = \epss(\omegaobs)$, but where no poles cross the real $k_x$-axis as we reduce $\im{\omega}$ from $\gamma$ to $0$, then the resulting field would be given by Fig. \ref{fig:wrong}b. This will not be the case for a conventional weak gain medium, such as that described by \eqref{inv_lorentz}, but will be possible for a sufficiently large $d$ using an advanced medium of the type considered in \cite{chen05,skaar06,nistad08}.

\appendix
\section{Absolute instabilities}\label{app:absinstab}
To have an absolute instability, poles of the field must collide in the complex $k_x$-plane, as we try to deform a certain segment of the inverse Laplace integration contour. This means $E(k_x,z,\omega)$ must have a pole of order $n \ge 2$ in the complex $k_x$-plane for a frequency with $\im{\omega}>0$. Define 
\begin{equation}\label{Ddef}
 D = (\kz + \kzz)^2 - (\kz - \kzz)^2\exp(2i\kzz d),
\end{equation}
which is the denominator in \eqref{RST}. Now, for $E(k_x,z,\omega)$ to have a pole of order $n \ge 2$, the derivative $\diff{D}/\diff{k_x}$ must be zero, at a zero of $D$. By differentiation
\begin{align}\label{dD}
	\frac{\diff{D}}{\diff{k_x}} & = 2(\kz + \kzz)(\dot{\kz} + \dot{\kzz})\nonumber \\ & - 2(\kz - \kzz)(\dot{\kz} - \dot{\kzz})e^{2i\kzz d} \\  & - 2i\dot{\kzz}d (\kz-\kzz)^2 e^{2i\kzz d}. \nonumber
\end{align}
Here, $\dot{\kz},\dot{\kzz}$ mean derivatives of $\kz,\kzz$ wrt. $k_x$. From the expressions for $\kz$ and $\kzz$ we find $\dot{\kz} = -{k_x}/{\kz}$ and $\dot{\kzz} = -{k_x}/{\kzz}$. The derivative is evaluated at a zero ($D=0$), so we may simplify \eqref{dD} to
\begin{align}
	\frac{\diff{D}}{\diff{k_x}} = -2k_x \frac{(\kz + \kzz)^2}{\kz\kzz} (2 - i\kz d).
\end{align}

From \eqref{Ddef} we see that there are no zeros for $\kz+\kzz=0$. The only possibilities for a zero of order $n\ge 2$ in the complex $k_x$-plane, are therefore $k_x=0$ or $k_z=-2i/d$. The last option means that the field is amplified in a passive dielectric, which is unphysical for large times, when all transients have died out. 

\bibliography{paper.bbl}

\end{document}